\documentclass[aps,twocolumn,superscriptaddress,groupedaddress,nofootinbib,floatfix]{revtex4}  

\usepackage{graphicx}  
\usepackage{bm}     
\usepackage{amsmath,amssymb,amsfonts}
\usepackage[utf8]{inputenc}
\usepackage{xspace}
\usepackage{physics}

\usepackage{hyphenat}
\usepackage{color}
\usepackage{booktabs}

\usepackage{hyperref}
\hypersetup{
    colorlinks=true,
    allcolors=[rgb]{0.26,0.41,0.88},
}

\usepackage{newtx}

\newcommand{\keVnr}{\text{keV}_{\text{nr}}}

\newcommand{\MeV}{\,\mathrm{MeV}}
\newcommand{\GeV}{\,\mathrm{GeV}}
\newcommand{\TeV}{\,\mathrm{TeV}}
\newcommand{\keV}{\,\mathrm{keV}}
\newcommand{\s}{\,\mathrm{s}}

\usepackage{cleveref}

\begin{document}
\title{Fermionic Dark Matter Absorption and the High-Energy Event in LUX-ZEPLIN}

\author{Yuanchao Lou}
\email[]{yuanchao\_lou@nnu.edu.cn}
\affiliation{Department of Physics and Institute of Theoretical Physics, Nanjing Normal University, Nanjing, 210023, China}

\author{Chih-Ting Lu}
\email[]{ctlu@njnu.edu.cn}
\affiliation{Department of Physics and Institute of Theoretical Physics, Nanjing Normal University, Nanjing, 210023, China}

\begin{abstract}
The LUX-ZEPLIN (LZ) experiment has reported a single candidate event in the high-energy nuclear recoil window $248\pm32.5\ \mathrm{keV}_{\mathrm{nr}}$ with an exposure of $2.84\ \mathrm{ton}\cdot\mathrm{yr}$, while the low-energy spectrum remains consistent with background expectations. We demonstrate that this excess can be naturally explained by the neutral-current absorption of fermionic dark matter on xenon nuclei. For a dark matter mass $m_\chi \simeq 247\ \mathrm{MeV}$, the coherent absorption process produces a monoenergetic nuclear recoil at $E_R \simeq 248\ \mathrm{keV}_{\mathrm{nr}}$. At this momentum transfer, the absorption process enters the incoherent regime, where scattering off individual nucleons produces a broad recoil spectrum extending from $\sim 200\ \mathrm{keV}$ to $100.2\ \mathrm{MeV}$. We show that a single effective field theory coupling can simultaneously produce one event in the $248\pm32.5\ \mathrm{keV}_{\mathrm{nr}}$ window while remaining consistent with the non-observation of events in neighboring energy regions. The required single-nucleon absorption cross section is $\sigma_{\chi N}^{\mathrm{NC}} = 1.07\times10^{-46}\ \mathrm{cm}^2$, corresponding to an effective field theory scale $\Lambda \simeq 11.5\ \mathrm{TeV}$. However, a recasting analysis of KamLAND data on the neutron-emission channel $\chi+{}^{12}\mathrm{C} \to \nu + n + {}^{11}\mathrm{C}^*$ excludes this benchmark parameter space, establishing a significant tension between the LZ excess interpretation and existing constraints from large-volume scintillator detectors. We discuss the implications of this tension and prospects for resolving it with future dedicated high-energy analyses.
\end{abstract}

\maketitle
\section{Introduction}

Direct detection experiments for particle dark matter (DM) are entering an unprecedented era of sensitivity. Dual-phase liquid xenon time projection chambers (LXe-TPCs), such as LUX-ZEPLIN (LZ), PandaX-4T, and XENONnT, have pushed cross-section constraints on Weakly Interacting Massive Particles (WIMPs) down toward the irreducible neutrino fog \cite{LZ2024SI,PandaX:2022osq,XENONnT2023}. In the absence of low-energy elastic scattering signatures ($E_R \lesssim 100\ \text{keV}_{\text{nr}}$), significant attention has turned toward expanding the experimental energy window to probe non-standard interaction topologies at higher recoil energies ($E_R \sim \mathcal{O}(100)\ \text{keV}_{\text{nr}}$) \cite{Fitzpatrick2012,Bramante2016}.

Recently, the LZ Collaboration presented a new preliminary analysis extending the DM search to higher nuclear recoil energies, up to $270\ \text{keV}_{\text{nr}}$, using an exposure of $2.84\ \text{tonne}\cdot\text{yr}$ from $220$ live-days of operation \cite{LZ2026}. While the low-energy spectrum remains consistent with background expectations, a single high-energy candidate event, designated LZ230616, was observed at: $E_R^{\text{obs}} = 248 \pm 23_{\text{stat}} \pm 23_{\text{sys}}\ \text{keV}_{\text{nr}}$, consistent with an elastic nuclear recoil \cite{LZ2026}. The event was located at $26.4\ \text{cm}$ from the cathode and $26.9\ \text{cm}$ from the TPC wall, with primary and secondary scintillation signals $S1_c = 540.1\ \text{phd}$ and $\log_{10}(S2_c) = 3.98$ \cite{LZ2026}. The LZ collaboration's background model predicts only $0.0106 \pm 0.0008_{\text{sys}}$ events in the $500 < S1_c < 600$ region where the event was detected, yielding a $3.4\sigma$ local significance ($2.6\sigma$ global after accounting for 616 EFT models tested) \cite{LZ2026}. The event is also difficult to explain with other backgrounds: the expected MSSI background is $0.005 \pm 0.005$ events, accidental coincidences are expected at $2.7 \pm 0.6$ events with $99.5\%$ rejection efficiency, and the neutron veto efficiency is $92 \pm 4\%$ \cite{LZ2026}. While the LZ collaboration does not identify this as a statistically significant excess, the event's energy and topology motivate systematic investigation.

Explaining a localized nuclear recoil at $E_R \sim 250\ \text{keV}_{\text{nr}}$ poses formidable challenges for conventional DM mechanisms. Standard elastic WIMP scattering produces a differential recoil spectrum that falls exponentially with energy; scaling the cross section to yield a single event at $250\ \text{keV}_{\text{nr}}$ overpredicts low-energy rates by orders of magnitude, conflicting with LZ's stringent constraints \cite{LZ2024SI}. Velocity-dependent or inelastic DM frameworks can impart larger momentum transfers but typically require fine-tuned mass splittings or exotic velocity distributions \cite{Hochberg2018,Bringmann2019,Cappiello2019}. Bosonic DM absorption (e.g., dark photons or axion-like particles) transfers rest mass energy primarily into electronic recoils via atomic ionization, failing to naturally generate a pure nuclear recoil signal \cite{Pospelov2008,An2015}.

In this work, we show that neutral-current (NC) absorption of fermionic DM on nuclei~\cite{Dror:2019onn,Dror:2019dib} provides a minimal and viable explanation for the LZ candidate event: $\chi + (A,Z) \rightarrow \nu + (A,Z)$. This exothermic process converts the entire rest mass energy $m_\chi$ into a monochromatic nuclear recoil: $E_R^0\simeq m_\chi^2 /2 m_A$. For $m_\chi \simeq 247\ \text{MeV}/c^2$, the recoil peak falls naturally at $E_R^0 \simeq 248\ \text{keV}_{\text{nr}}$, matching the observed event. For $m_\chi \gtrsim 100\ \text{MeV}$, the momentum transfer $q \simeq m_\chi \sim p_F^{\text{Xe}} \simeq 262\ \text{MeV}$ resolves individual nucleons, giving rise to an inevitable incoherent single-nucleon knockout channel alongside coherent nuclear recoils \cite{Ge:2024euk}. This correlated coherent-incoherent signal provides a distinctive signature of fermionic absorption.

We find that a single EFT coupling with $\sigma_{\chi N}^{\text{NC}} = 1.07\times10^{-46}\ \text{cm}^2$ ($\Lambda \simeq 11.5\ \text{TeV}$) reproduces the observed event via coherent absorption, while the predicted $\mathcal{O}(10^2)$ primary incoherent knockouts are expected to be rejected by standard LZ selection cuts, ensuring consistency with the clean low-energy spectrum. We compute the full coherent and incoherent spectra, discuss detector-level implications, and highlight a significant tension between the LZ excess interpretation and existing constraints
from large-volume scintillator detectors, such as KamLAND~\cite{KamLAND:2011bnd} and JUNO~\cite{JUNO:2022lpc}.

The paper is organized as follows. Sec.~\ref{sec:fermion_model} introduces the EFT framework and UV completions. Sec.~\ref{sec:results} shows that coherent absorption can explain the LZ event and predicts an associated incoherent component. Sec.~\ref{sec:KamLAND} presents a KamLAND recasting on the neutron‑emission channel, which strongly disfavors the required parameter space. Sec.~\ref{sec:constraints} summarizes other constraints, and Sec.~\ref{sec:conclusions} concludes. Details are given in the Supplemental Material.

\section{The models}
\label{sec:fermion_model}

\textbf{Effective Field Theory:} Fermionic dark matter (FDM) absorption by nuclei, $\chi+(A,Z)\to\nu_R+(A,Z)$, provides a distinctive signal for sub-GeV dark sectors. At the nucleon level $N\in\{n,p\}$, the interaction is described by dimension-six four-fermion operators~\cite{Dror:2019onn,Dror:2019dib}:
\begin{align}
\mathcal{O}_{\rm NC}^V &= \frac{1}{\Lambda_V^2} (\bar{n}\gamma^\mu n+\bar{p}\gamma^\mu p)(\bar{\chi}\gamma_\mu P_R\nu_R)+\mathrm{H.c.}, \label{eq:V} \\
\mathcal{O}_{\rm NC}^S &= \frac{1}{\Lambda_S^2}\sum_{N=n,p}g_S^N(\bar{N}N)(\bar{\chi}P_R\nu_R)+\mathrm{H.c.}, \label{eq:S}
\end{align}
where $P_R=(1+\gamma_5)/2$ and $\nu_R$ is a light right-handed Dirac neutrino. For a heavy target, energy conservation gives $E_\nu\sim m_\chi$ and $E_R\sim m_\chi^2/(2m_A)$; for xenon, $m_\chi=247\MeV$ yields $E_R\simeq248\keV$, matching the LZ high-energy recoil candidate. The benchmark mass lies in the interval, $m_{\pi^0} < m_\chi < 2m_{\pi^0}$, which permits single-pion anomaly processes while forbidding two-pion hadronic cuts. The absorption cross section per nucleon is $\sigma_{\chi N}^{\rm NC}\simeq m_\chi^2/(4\pi\Lambda^4)$ for the vector operator, while the scalar operator scales as $g_S^pZ+g_S^n(A-Z)$, recovering $A^2$ enhancement only in the isoscalar limit.

\textbf{Vector Portal (Gauged $U(1)_B$):} To avoid tree-level $\chi\to\nu_R e^+e^-, \nu_R \mu^+\mu^-$ decays, we gauge a purely leptophobic $U(1)_B$ with $G_V=G_{\rm SM}\times U(1)_B$. SM quarks carry $B=1/3$; leptons carry $B=0$. Anomaly cancellation requires a spectator sector of vector-like fermions ($\Psi,\eta,\xi$) with charges arranged to cancel all mixed and cubic anomalies; their masses are generated by a singlet $\zeta\sim(1,1,0,3)$ after $U(1)_B$ breaking. DM is a vector-like Dirac fermion $\chi\sim(1,1,0,1)$, and $\nu_R\sim(1,1,0,0)$. A scalar $\Phi_B\sim(1,1,0,1)$ with $\vev{\Phi_B}=v_B/\sqrt2$ generates the renormalizable mixing $\mathcal{L}_{\chi\nu}^V=-m_{\chi0}\bar{\chi}_L\chi_R-y_{\chi\nu}\Phi_B\bar{\chi}_L\nu_R+\mathrm{H.c.}$. After symmetry breaking, $m_\chi=\sqrt{m_{\chi0}^2+\delta^2}$ with $\delta=y_{\chi\nu}v_B/\sqrt2$, and the right-handed mixing angle satisfies $\sin\theta\equiv s_\theta=\delta/m_\chi$, $\cos\theta\equiv c_\theta=m_{\chi0}/m_\chi$. Integrating out the $U(1)_B$ gauge boson $V$ of mass $m_V$ yields
\begin{equation}
\frac{1}{\Lambda_V^2}=\frac{g_B^2s_\theta c_\theta}{m_V^2},
\end{equation}
with benchmark $g_B=0.30,\ s_\theta=0.50$ giving $m_V\simeq2.27\TeV$ for $\Lambda_V=11.5\TeV$. Kinetic mixing $\epsilon_B V_{\mu\nu}B_Y^{\mu\nu}$ must be suppressed to avoid leptonic decay modes. In the special mass window, the Wess–Zumino–Witten anomaly induces $\chi\to\nu_R\pi^0\gamma$ with width, requiring a complete anomaly-matching calculation. See Supplemental Material for more details about the UV completion.

\textbf{Scalar Portal (Heavy Gluonic Mediator):} We introduce a heavy CP-even singlet mediator $\phi$ with $m_\phi\gg m_\chi$ to avoid rapid $\chi\to\nu_R\phi$ and coupling $\mathcal{L}_{\chi\nu,S}=-y_{\chi\nu}\phi\bar{\chi}P_R\nu_R+\mathrm{H.c.}$. The nucleon coupling is generated through a heavy colored sector matching to
\begin{equation}
\mathcal{L}_{\phi g}=\frac{\alpha_s}{12\pi}\frac{\kappa_g}{v}\phi G_{\mu\nu}^aG^{a\mu\nu}.
\end{equation}
Using the QCD trace anomaly, $\left\langle N|\alpha_s G^2/\pi|N\right\rangle=-\frac{8}{9}m_N f_{TG}^N\bar NN$, with $f_{TG}^N=1-\sum_{q=u,d,s}f_{Tq}^N$, we find
\begin{equation}
g_S^N=-\frac{m_N}{18v}f_{TG}^N,\qquad \frac{1}{\Lambda_S^2}=\frac{y_{\chi\nu}\kappa_g}{m_\phi^2}.
\end{equation}
For $m_\phi=20\GeV,\ \kappa_g=0.1$, $\Lambda_S=11.5\TeV$ gives $y_{\chi\nu}\simeq3\times10^{-5}$. Since $m_\phi>m_K-m_\pi\simeq354\MeV$, the on-shell $K^+\to\pi^+\phi$ is forbidden. The dominant decay is $\chi\to\nu_R\gamma\gamma$; because $m_\chi<2m_{\pi^0}$, the free-quark optical-theorem argument is invalid, and a dispersive hadronic calculation is required. See Supplemental Material for more details about the UV completion.

\textbf{Relic Abundance:} In both portals, the same mediator that generates absorption would be difficult to play the crucial role in DM annihilation for fitting the observed relic abundance~\cite{Cox:2023cjw}. We decouple the relic sector via a separate CP-odd singlet $a$ with $m_a>m_\chi$ and interaction $\mathcal{L}_{a\chi}=-ig_a a\bar{\chi}\gamma^5\chi$. Forbidden annihilation $\chi\bar\chi\to aa$ on the thermal tail fixes $\Omega_\chi h^2\simeq0.12$. The pseudoscalar $a$ decays before BBN through small Higgs-portal mixing. More details are shown in Supplemental Material.

\begin{table}[h]
\centering
\caption{Summary of vector and scalar UV portals.}
\label{tab:models}
\begin{tabular}{lcc}
\toprule
Feature & Vector $U(1)_B$ & Scalar Gluonic \\
\midrule
Mediator & $V$ (gauge boson) & $\phi$ (heavy scalar) \\
Matching & $1/\Lambda_V^2=g_B^2s_\theta c_\theta/m_V^2$ & $1/\Lambda_S^2=y_{\chi\nu}\kappa_g/m_\phi^2$ \\
Nucleon coupling & $A^2$ coherent & $g_S^N A^2$ (isoscalar) \\
Relic mechanism & Forbidden $\chi\bar\chi\to aa$ & Forbidden $\chi\bar\chi\to aa$ \\
Leading decay & $\chi\to\nu_R\pi^0\gamma$ (WZW) & $\chi\to\nu_R\gamma\gamma$ \\
Special mass window & WZW active & $\chi\to\nu_R\pi\pi$ closed \\
\bottomrule
\end{tabular}
\end{table}

Table I summarizes the two frameworks; complete details are provided in the Supplemental Material.

\section{FDM Absorption Meets LZ230616}
\label{sec:results}

\subsection{Coherent absorption and the LZ230616 event}

The LUX-ZEPLIN (LZ) experiment, with an exposure of $2.84\ \text{tonne}\cdot\text{yr}$ from $220$ live-days of operation, has reported a single nuclear-recoil candidate event in the high-energy window \cite{LZ2026}. The event, designated LZ230616, occurred on 16 June 2023 and was reconstructed with energy $248\pm23_{\rm stat}\pm23_{\rm sys}\ \keVnr$, consistent with an elastic nuclear recoil \cite{LZ2026}. The low-energy nuclear recoil spectrum ($E_R\lesssim100\ \keVnr$) remains consistent with background expectations. The LZ collaboration's official analysis finds no statistically significant excess over expected backgrounds, reporting a local significance of $3.4\sigma$ and a global significance of $2.6\sigma$ across 616 EFT models tested \cite{LZ2026}. Nevertheless, the event is noteworthy: the expected MSSI background in the science sample is only $0.005\pm0.005$ events; accidental coincidences are expected at $2.7\pm0.6$ events with $99.5\%$ rejection efficiency; the neutron veto efficiency is $92\pm4\%$; and the combined background expectation in the $500<S1c<600$ region is $0.0106\pm0.0008$ events \cite{LZ2026}. These considerations make the event difficult to explain with known Standard Model backgrounds, motivating systematic investigation.

The absorption of FDM on nuclear targets, $\chi+(A,Z)\to\nu+(A,Z)$, was first experimentally searched by the PandaX-4T collaboration~\cite{PandaX:2022osq}. Using a $0.63\ \text{tonne}\cdot\text{yr}$ exposure, PandaX-4T set a $90\%$ C.L. upper limit of $1.5\times10^{-50}\ \text{cm}^2$ at $m_\chi=40\ \text{MeV}$. The MAJORANA Demonstrator subsequently searched for fermionic absorption using $^{76}\text{Ge}$ detectors in a $37.5\ \text{kg}\cdot\text{yr}$ exposure \cite{Majorana:2022gtu}. More recently, PICO-60 performed a dedicated search in a C$_3$F$_8$ bubble chamber, setting leading constraints on spin-independent absorption below approximately $23\ \text{MeV}$ and first limits on spin-dependent absorptive interactions \cite{PICO:2025rku}.

For FDM masses below $\sim100\ \text{MeV}$, the absorption process is dominated by coherent scattering off the entire nucleus, with the rate scaling as $A^2F^2(q^2)$ where $F(q)$ is the form factor. However, for $m_\chi\gtrsim100\ \text{MeV}$, the momentum transfer $q\simeq m_\chi$ becomes large enough to resolve individual nucleons, and the incoherent scattering regime becomes significant \cite{Ge:2024euk}. For $m_\chi\sim250\ \text{MeV}$, $q\sim p_F$ where the Fermi momentum $p_F\simeq262\ \text{MeV}$ for xenon, placing this system squarely in the coherent-to-incoherent transition regime.

The differential rate for coherent nuclear recoil absorption is given by \cite{PandaX:2022osq,Ge:2024euk}:
\begin{equation}
\frac{dR_{\text{coh}}}{dE_R}
=
\frac{\rho_\chi}{m_\chi}
\frac{\sigma_{\chi N}^{\text{NC}}}{M_T}
\sum_j N_j M_j A_j^2 F_{H,j}^2(q_j)
\frac{q_j}{p_{\nu,j}}
\left\langle \frac{1}{v} \right\rangle_{v>v_{\min,j}},
\end{equation}
where $\rho_\chi=0.3\ \text{GeV}/\text{cm}^3$ is the local DM density, $\sigma_{\chi N}^{\text{NC}}\simeq m_\chi^2/(4\pi\Lambda^4)$ is the absorption cross section per nucleon, $M_T$ is the total target mass, $N_j$ and $M_j$ are the number and mass of isotope $j$, $A_j$ is the mass number, $F_{H,j}(q_j)$ is the Helm form factor shown in the Supplemental Material, $q_j=\sqrt{2M_jE_R}$, $p_{\nu,j}=\sqrt{q_j(2m_\chi-q_j-2E_R)}$, and $\langle1/v\rangle_{v>v_{\min,j}}$ is the velocity-averaged inverse speed from the Standard Halo Model \cite{Dror:2019dib}.

For $^{131}\text{Xe}$ at $m_\chi=247\ \text{MeV}$, $q\simeq m_\chi$, and the coherent enhancement factor using the Helm parametrization is:
\begin{equation}
\mathcal{C}_{\text{coh}}=A^2F_H^2(q^2)\simeq2.4,
\end{equation}
which should be understood as a \emph{Helm-model coherent-response benchmark}, as $q\sim250\ \text{MeV}$ is very close to the nuclear-structure-sensitive regime. Different form-factor parameterizations (Helm, Klein-Nystrand, symmetrized Fermi) may differ by up to an order of magnitude near dips \cite{Ge:2024euk}.

\begin{figure}[h]
\centering
\includegraphics[width=0.85\columnwidth]{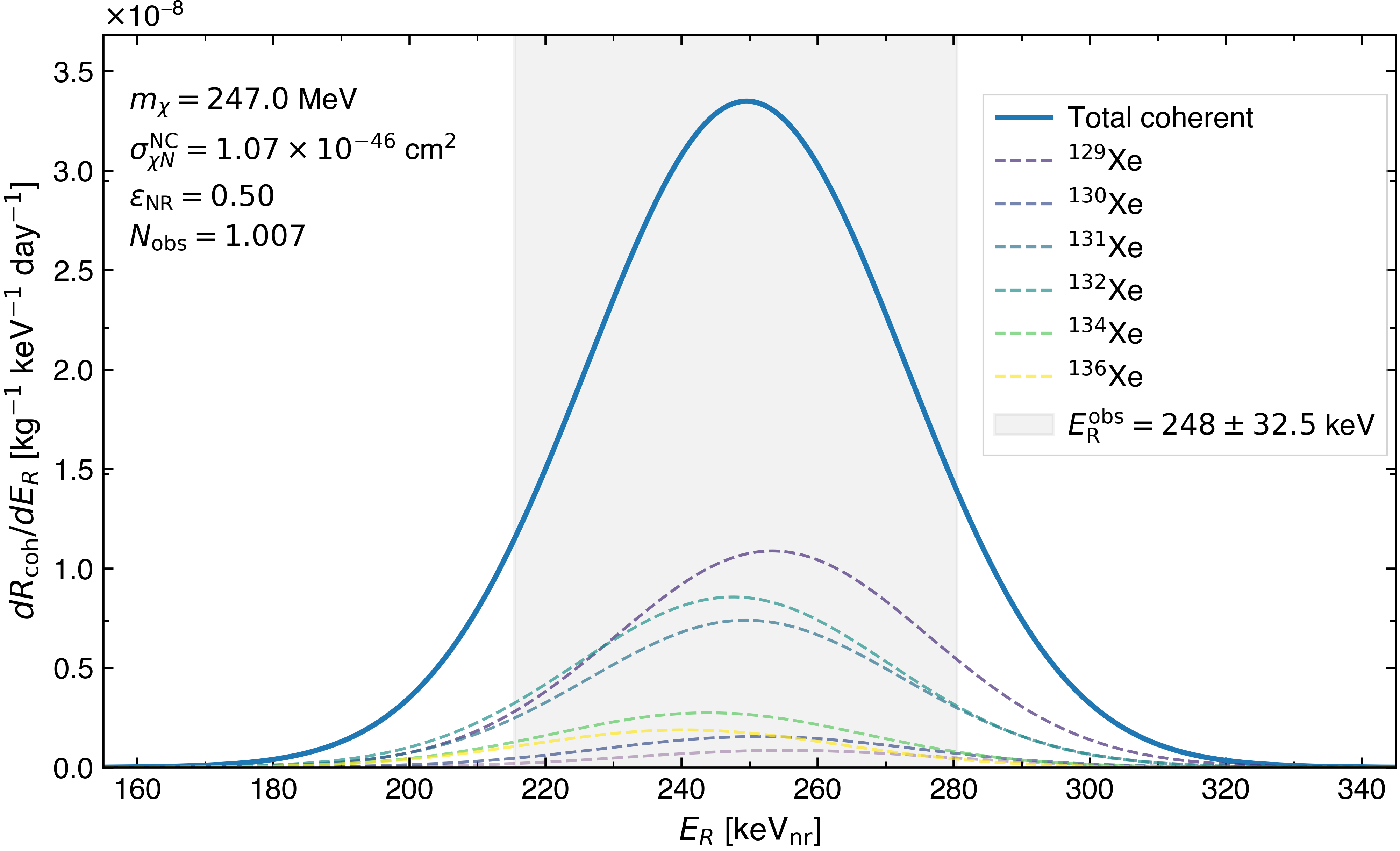}
\caption{Coherent fermionic dark-matter absorption spectrum for the benchmark $m_\chi=247\ \text{MeV}$ and $\sigma_{\chi N}^{\rm NC}=1.07\times10^{-46}\ \text{cm}^2$, with exposure $2.84\ \text{tonne}\cdot\text{yr}$, $220$ live-days, and $\sigma_E=23\ \keVnr$. The dashed curves show the dominant xenon isotopes and the solid curve their sum; the shaded band denotes the $248\pm32.5\ \keVnr$ ROI. The apparent width is dominated by the detector energy resolution.}
\label{fig:coh}
\end{figure}

Figure~\ref{fig:coh} shows the coherent differential spectrum for the benchmark parameters $m_\chi=247\ \text{MeV}$ and $\sigma_{\chi N}^{\text{NC}}=1.07\times10^{-46}\ \text{cm}^2$ (corresponding to $\Lambda\simeq11.5\ \text{TeV}$), convolved with a Gaussian energy resolution of $\sigma_E=23\ \text{keV}$. The detection efficiency in the ROI is approximately $0.5$ \cite{LZ2026}. The spectrum consists of a series of isotope-dependent peaks centered at:
\begin{equation}
E_{R,j}^0=\frac{m_\chi^2}{2(M_j+m_\chi)},
\end{equation}
which for the dominant xenon isotopes lie between $246.4\ \text{keV}$ (${}^{136}\text{Xe}$) and $260.8\ \text{keV}$ (${}^{128}\text{Xe}$), all within the $248\pm32.5\ \text{keV}$ window. The peak width is dominated by the detector energy resolution ($\sim23\ \text{keV}$) rather than the physical process itself, reflecting the LZ liquid xenon detector response function.

Using the benchmark cross section, the expected number of coherent events in the $215.5$--$280.5\ \text{keV}$ window is:
\begin{equation}
N_{\text{coh}}=\int_{215.5}^{280.5}
\frac{dR_{\text{coh}}}{dE_R}
\,dE_R
\times\mathcal{E}
\simeq1.0,
\end{equation}
where $\mathcal{E}=2.84\ \text{tonne}\cdot\text{yr}$ is the exposure with $220$ live-days. This matches the observed single event. The $248\pm32.5\ \text{keV}$ feature is naturally the coherent nuclear recoil channel, not the incoherent nucleon recoil channel. In this energy window, incoherent scattering contributes negligibly because the kinematics of single-nucleon knockout are phase-space suppressed for keV-scale xenon recoils. For $E_R<200\ \text{keV}$, the incoherent contribution is also negligible: the minimum outgoing nucleon kinetic energy allowed by Pauli blocking is $T_{N'}^{\min}\simeq0.2\ \text{MeV}$ (PWIA) or $\sim31$--$39\ \text{MeV}$ (Hard RFG), corresponding to xenon recoil energies far below the experimental threshold~\cite{Ge:2024euk}. Here, PWIA and RFG refer to the Plane Wave Impulse Approximation and the Relativistic Fermi Gas, respectively. Thus, the incoherent signal cannot account for any observable low-energy events, explaining why LZ observes background consistency at low energies while seeing a single event at $250\ \text{keV}$.

Crucially, $q\simeq250\ \text{MeV}\sim p_F^{\text{Xe}}\simeq262\ \text{MeV}$, placing this interaction in the core transition regime between coherent nuclear scattering and incoherent single-nucleon knockout. This is the key physical insight: the LZ high-energy feature provides a probe of the transition from coherent to incoherent FDM absorption in a heavy nucleus.

\subsection{The unavoidable incoherent component}

The incoherent process $\chi+N_{\text{bound}}\to\nu+N'$ is characterized by the outgoing-nucleon kinetic energy $T_{N'}=E_{N'}-m_N$, which is not identical to the reconstructed recoil energy of the entire xenon nucleus. The differential production rate is \cite{Ge:2024euk}:
\begin{equation}
\frac{dR_{\text{incoh}}}{dT_{N'}}
=
\frac{\rho_\chi}{m_\chi}
\frac{1}{M_T}
\sum_j N_j
\left[
Z_j \frac{d\langle \sigma_p v_p \rangle}{dT_p}
+ (A_j-Z_j) \frac{d\langle \sigma_n v_n \rangle}{dT_n}
\right], 
\end{equation}
where $d\langle \sigma_N v_N \rangle/dT_N$ is the Fermi-averaged differential cross section for a single nucleon. For the benchmark, the allowed primary-nucleon kinetic energy spans approximately $0.2$--$100.2\ \text{MeV}$.

\begin{figure}[h]
\centering
\includegraphics[width=0.85\columnwidth]{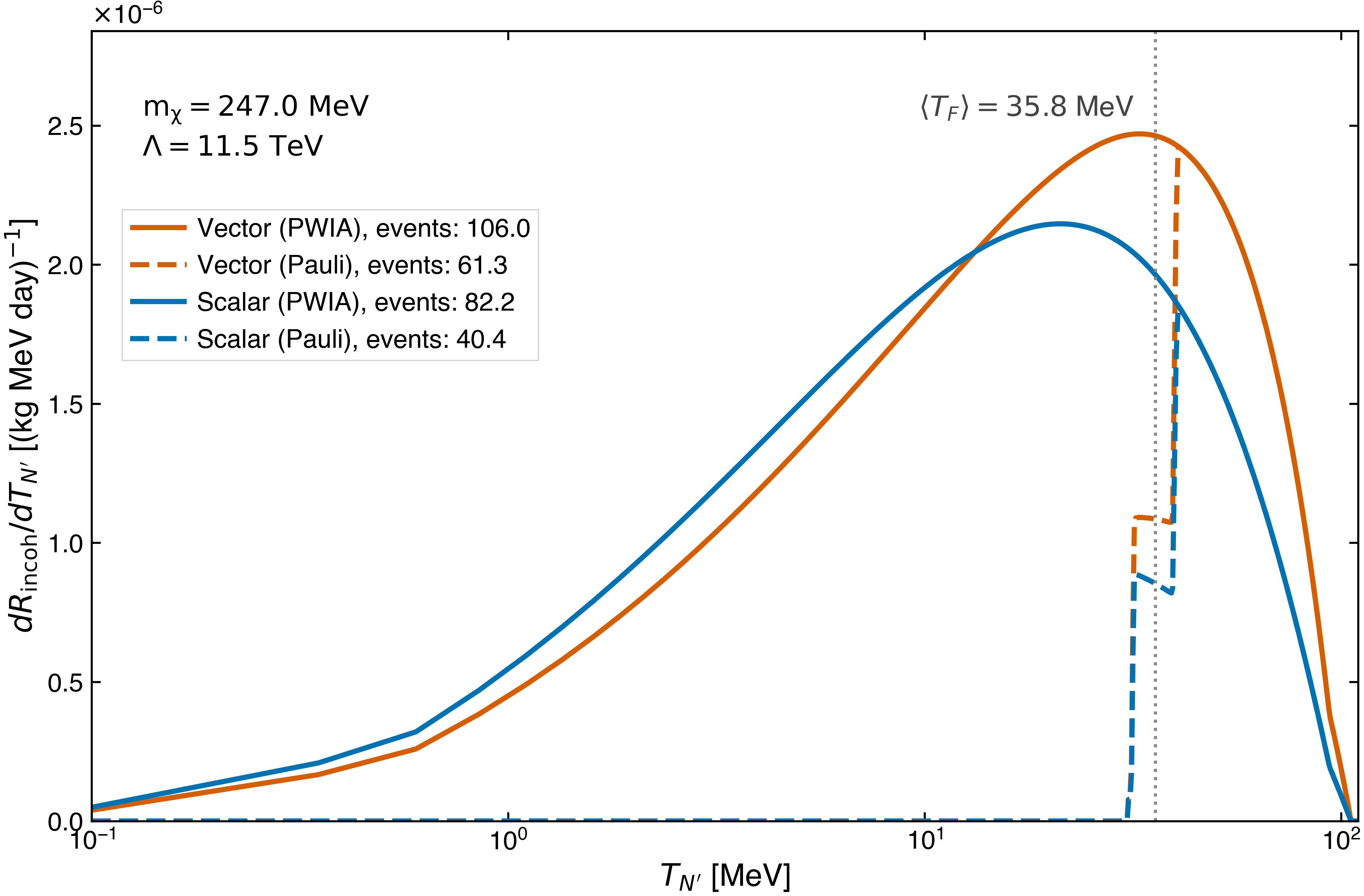}
\caption{Incoherent absorption spectra as a function of the outgoing-nucleon kinetic energy $T_{N'}$. Solid curves correspond to the PWIA/form-factor prescription, while dashed curves include hard RFG Pauli blocking. The vertical line marks the representative xenon Fermi-energy scale $\langle T_F\rangle\simeq35.8\ \text{MeV}$. The scalar interaction is suppressed relative to the vector interaction due to its momentum-dependent matrix element.}
\label{fig:incoh}
\end{figure}

Figure~\ref{fig:incoh} shows the incoherent spectrum computed with exact matrix elements (see Supplemental Material), using both the PWIA/Form-Factor Prescription and the hard RFG Pauli Blocking methods.

\textbf{PWIA/Form-Factor Prescription:} In the Plane Wave Impulse Approximation (PWIA), the outgoing nucleon is treated as a free particle, and the nuclear response is described by the form factor $1-F^2(q^2)$. This approach allows the final-state nucleon to occupy any momentum state, including those below the Fermi surface, and thus provides an upper bound on the incoherent rate~\cite{Ge:2024euk}.

\textbf{Hard RFG Pauli Blocking:} In this approach, the final-state nucleon is required to have momentum $|\mathbf{p}_{N'}|>p_F$, implementing the Pauli exclusion principle exactly within the Relativistic Fermi Gas (RFG) model. This yields a lower bound on the incoherent rate, as low-lying final states are forbidden~\cite{Ge:2024euk}. The completely degenerate Fermi-gas Pauli-blocking approximation overestimates blocking; using the nuclear form factor provides a more realistic incoherent estimate~\cite{Ge:2024euk}. 

\begin{table}[h]
\centering
\caption{Total incoherent primary production events for the benchmark exposure of $2.84\ \text{tonne}\cdot\text{yr}$ with $220$ live-days. These are theoretical production rates; reconstructed LZ rates require additional topology-dependent efficiency.}
\label{tab:incoh}
\begin{tabular}{lcc}
\toprule
Method & Vector & Scalar \\
\midrule
PWIA/Form-Factor & 106.05 & 82.23 \\
Hard RFG Pauli & 61.26 & 40.44 \\
\bottomrule
\end{tabular}
\end{table}

The numerical results are summarized in Table~II. The scalar-to-vector ratios are $0.775$ (PWIA) and $0.660$ (Pauli). This suppression arises because the scalar matrix element contains an explicit momentum transfer factor that penalizes high relativistic momentum transfers relative to the vector interaction (see Supplemental Material).

\textbf{Absence of a sharp cutoff at $35.8\ \text{MeV}$:} The average Fermi energy of xenon is $\langle T_F\rangle\simeq35.8\ \text{MeV}$ ($p_{F,\text{avg}}\simeq0.260\ \text{GeV}$). However, the Pauli-blocked curves do not exhibit a single sharp cutoff at $35.8\ \text{MeV}$. This is due to isospin separation: xenon has an excess of neutrons ($Z=54,\ A-Z=77$), yielding distinct Fermi momenta: $p_{F,p}=p_{F,\text{avg}}\left(\frac{2Z}{A}\right)^{1/3}\simeq0.244\ \text{GeV}$ and $p_{F,n}=p_{F,\text{avg}}\left(\frac{2(A-Z)}{A}\right)^{1/3}\simeq0.274\ \text{GeV}$ such that $T_{F,p}\simeq31.2\ \text{MeV}$ and $T_{F,n}\simeq39.2\ \text{MeV}$. Below $31.2\ \text{MeV}$, both proton and neutron knockouts are Pauli-forbidden. Between $31.2\ \text{MeV}$ and $39.2\ \text{MeV}$, the proton channel unblocks while the neutron channel remains blocked, producing the two-step feature. Above $39.2\ \text{MeV}$, both channels are fully open and merge with the PWIA curves.

\textbf{High-energy inflection point at $T_{N'}\simeq93.4\ \text{MeV}$:} The inflection point marks the maximal kinematic energy limit for head-on collisions with protons ($p_{F,p}\simeq0.244\ \text{GeV}$). Above $93.4\ \text{MeV}$, the proton phase space closes entirely; only the neutron channel continues to contribute up to its absolute kinematic ceiling at $T_{N',\max,n}\simeq104.8\ \text{MeV}$, where the spectrum vanishes.

\subsection{Detector-level implications and the coherent-incoherent correlation}

For xenon recoil energies above $300\ \text{keV}$, the coherent signal is negligible. However, the incoherent signal extends from approximately $0.2\ \text{MeV}$ to $100.2\ \text{MeV}$ in $T_{N'}$ and the bulk of the incoherent signal lies in the $T_{N'}\sim10$--$80\ \text{MeV}$ range. The LZ experiment is designed and calibrated for low-energy nuclear recoils up to $\sim100\ \keVnr$ \cite{LZ2026}. For recoils above $300\ \text{keV}$, several factors affect detection: (i) the energy resolution for high-energy nuclear recoils has not been characterized; (ii) standard WIMP selection cuts may reject high-energy events; (iii) the nuclear recoil response model is calibrated only up to $\sim80\ \keVnr$ \cite{LZ2026}. Under standard LZ low-energy WIMP search selections, the recorded incoherent event count is 0, as these events are filtered out by quality and energy cuts.

Crucially, LZ's veto systems provide additional suppression of high-energy events \cite{LZ2026}. The prompt veto tags events with Skin signal $>2.5\ \text{phd}$ or Outer Detector signal $>4.5\ \text{phd}$ within $\pm0.25$--$0.3\ \mu\text{s}$ of S1, with tagging efficiencies of $94\pm2\%$ for MSSI and $88\pm2\%$ for $^{127}\text{Xe}$ and detector gamma rays. The delayed veto tags events with Skin signal $>300\ \text{keV}$ or Outer Detector signal $>200\ \text{keV}$ within $600\ \mu\text{s}$ of S1, with neutron tagging efficiency of $87\pm2\%$ \cite{LZ2026}. These vetoes strongly suppress events originating from high-energy processes such as the incoherent nucleon knockouts predicted by the FDM absorption model.

Given the current lack of published LZ efficiency for nuclear recoils above $300\ \text{keV}$, we cannot determine how many incoherent events would be observed. However, we can estimate the total incoherent event rate that would be generated theoretically: $N_{\text{incoh}}^{\text{total}}\simeq106.05\ \text{events} \quad (\text{vector, PWIA})$, and $N_{\text{incoh}}^{\text{total}}\simeq61.26\ \text{events} \quad (\text{vector, Pauli})$. 
The true observable number would be $N_{\text{obs}}=\epsilon_{\text{LZ}}(T_{N'})\times N_{\text{incoh}}$, where $\epsilon_{\text{LZ}}(T_{N'})$ is the LZ detection efficiency for high-energy nuclear recoils/cascades. This efficiency is currently unknown. The predicted $O(10^2)$ primary incoherent knockouts are expected to be rejected by standard LZ selection cuts and veto systems, establishing a testable coherent-incoherent correlation for future dedicated high-energy analyses.

Future dedicated analyses targeting high-energy nuclear recoils could significantly improve sensitivity. Complementary experiments such as KamLAND~\cite{KamLAND:2011bnd} and JUNO~\cite{JUNO:2022lpc}, with their large exposures and sensitivity to higher energies, could provide independent probes. The characteristic broad spectrum of incoherent absorption, combined with the fixed ratio of coherent to incoherent rates, provides a unique signature that can be tested in future searches. A distinctive coherent absorption line embedded in a calculable incoherent nuclear-response background would provide strong evidence for the fermionic absorption mechanism.

\section{The tension from KamLAND recasting}
\label{sec:KamLAND}

The incoherent knockouts predicted in Sec.~\ref{sec:results} produce neutrons with kinetic energies in the $10$--$100\ \text{MeV}$ range, accessible to large‑volume scintillators via prompt‑delayed coincidences. KamLAND, with its large exposure, provides a strong test for the incoherent absorption. We recast KamLAND data to constrain the benchmark parameters and find a significant tension with the LZ interpretation.

\textbf{Signal signature:} KamLAND provides an additional test of the incoherent absorption on carbon.  In a liquid scintillator, the neutron emission channel
\begin{equation}
\chi+{}^{12}\mathrm{C}
\rightarrow
\nu+n+{}^{11}\mathrm{C}^{\ast}
\label{eq:kamland-process}
\end{equation}
can produce a prompt--delayed coincidence.  The prompt signal is
generated primarily by elastic scattering of the knocked-out neutron on
protons in the scintillator, $n+p\rightarrow n+p$,
where the recoil protons produce scintillation light~\cite{Gong:2025ves}. After
thermalization, the neutron is captured on hydrogen, $n+p\rightarrow d+\gamma(2.2\MeV)$, providing the delayed signal~\cite{Gong:2025ves,KamLAND:2021gvi}.

\textbf{Signal event estimation:} We calculate the prompt visible-energy spectrum induced by the
neutron-emission process in Eq.~\eqref{eq:kamland-process}.  Before
applying the KamLAND event-selection efficiency, the spectrum is~\cite{Ge:2024euk}
\begin{align}
\frac{dN}{dE_{\rm prompt}}
={}&
\mathcal N_{\rm C}^{(s)}
\frac{\rho_\chi}{m_\chi}c\,\sigma_0 N_n
\int_0^\infty 4\pi p^2dp
\int dE_{\rm miss}
\mathcal S_n(p,E_{\rm miss})
\nonumber\\
&\times
\int dT_n
\mathcal K_{\mathcal O}
(m_\chi,p,E_{\rm miss},T_n)
\delta\!\left[
E_{\rm prompt}-E_{\rm vis}(T_n)
\right], 
\label{eq:kamland-prompt-rate}
\end{align}
where $c$ is the speed of light. For the $6.72~\mathrm{kt\,yr}$ KamLAND exposure, we use
$\mathcal N_{\rm C}^{(s)}=9.09\times10^{39}~\mathrm{s}$.
The spectral function $\mathcal S_n(p,E_{\rm miss})$ describes the
momentum and missing energy distribution of an initial neutron in
${}^{12}\mathrm C$ and is normalized per neutron~\cite{Benhar:1994hw}.  The factor $N_n=6$
sums the response over the six neutrons in each carbon nucleus.

For each initial state $(p,E_{\rm miss})$, the outgoing neutron
kinematics are obtained by imposing energy conservation.  Defining
\begin{equation}
E_n^\prime=m_n+T_n,
\quad
p_n^\prime=\sqrt{(E_n^\prime)^2-m_n^2},
\end{equation}
the outgoing neutrino energy and the allowed angle are
\begin{equation}
E_\nu=m_\chi+m_n-E_{\rm miss}-E_n^\prime,
\qquad
\mu_\ast\equiv\cos\theta_\ast
=
\frac{p^2+(p_n^\prime)^2-E_\nu^2}
{2p\,p_n^\prime}.
\end{equation}
We retain only events with $E_\nu>0$, $|\mu_\ast|\leq1$, and
$p_n^\prime>p_F^{\rm C}$, where
$p_F^{\rm C}=220.5~\mathrm{MeV}$ is the carbon Fermi momentum used in
the hard Pauli-blocking prescription~\cite{Gong:2025ves}.

Following the absorption formalism of Ref.~\cite{Ge:2024euk}, the kernel
$\mathcal K_{\mathcal O}$ contains the elementary absorption matrix
element, where $\mathcal O=V,S$ labels the vector or scalar operator.
For the two operators used in the numerical calculation, respectively,
\begin{equation}
\mathcal K_V=
\frac{B_V}{2m_\chi^3E_p p},
\qquad
\mathcal K_S=
\frac{g_{S,n}^2 B_S}{4m_\chi^3E_p p},
\end{equation}
where $E_p=\sqrt{m_n^2+p^2}$ is the on-shell initial neutron energy in
the matrix element and
\begin{align}
B_V &=
(p^\prime\!\cdot k^\prime)(p\!\cdot k)
+(p^\prime\!\cdot k)(p\!\cdot k^\prime)
-m_n^2(k\!\cdot k^\prime),
\\
B_S &=
(p\!\cdot p^\prime+m_n^2)(k\!\cdot k^\prime).
\end{align}

The outgoing neutron deposits energy through scattering on protons in
the scintillator.  We treat $T_n^{\rm eff}$ as the effective energy
available for proton recoils after the adopted FSI energy-loss
prescription~\cite{Bodek:2019,Gong:2025ves}.  The prompt energy is calculated with Birks quenching~\cite{Birks:1951boa,Chou:1952},
\begin{equation}
E_{\rm prompt}=E_{\rm vis}(T_n)
=
\int_0^{T_n^{\rm eff}}
\frac{dT}
{1+k_B S_p(T)+C S_p^2(T)} ,
\label{eq:kamland-birks}
\end{equation}
where $T$ is the proton kinetic energy and $S_p(T)$ is the proton
stopping power in the KamLAND scintillator~\cite{Berger:1999}.  We use $k_B=7.79\times10^{-3}~
\mathrm{g\,cm^{-2}\,MeV^{-1}}$, and $C=1.64\times10^{-5}~
\bigl(\mathrm{g\,cm^{-2}\,MeV^{-1}}\bigr)^2$~\cite{Yoshida:2010zzb}.
The delta function in Eq.~\eqref{eq:kamland-prompt-rate} assigns each
outgoing-neutron energy $T_n$ to the corresponding prompt energy
$E_{\rm prompt}$.  This is an effective response model, rather than a
full event-by-event simulation of neutron transport and detector
response.

For a prompt-energy bin $[E_i,E_{i+1}]$, the selected signal prediction is
\begin{equation}
s_i
=
\epsilon_{\rm KL}
\int_{E_i}^{E_{i+1}}
dE_{\rm prompt}\,
\frac{dN}{dE_{\rm prompt}},
\label{eq:kamland-bin-signal}
\end{equation}
where we use $\epsilon_{\rm KL}=0.58$ and eight $3~\mathrm{MeV}$ bins
over $7.8<E_{\rm prompt}<31.8~\mathrm{MeV}$, following the KamLAND
recast and data treatment in
Refs.~\cite{Gong:2025ves,MeighenBerger:2023tad,KamLAND:2021gvi}.

\textbf{Exclusion limit:} Using the binned prompt delayed KamLAND data,
we derive preliminary $90\%$ confidence level lower bounds on the
interaction scales for both operators. At
$m_\chi \simeq 247~\mathrm{MeV}$, the present KamLAND recasting gives
\begin{equation}
\Lambda_{90}^{V}(\text{KamLAND})\gtrsim 80~\mathrm{TeV},
\qquad
\Lambda_{90}^{S}(\text{KamLAND})\gtrsim 73~\mathrm{TeV}.
\end{equation}
Both bounds are substantially larger than the scale
$\Lambda\simeq11.5~\mathrm{TeV}$ required by the LZ benchmark.
Thus, within the present signal and detector response treatment, the
KamLAND recasting places both the vector and scalar fermionic
absorption interpretations of the LZ excess in tension. The binned
likelihood construction and statistical procedure are given in the
Supplemental Material.

\section{Other constraints}
\label{sec:constraints}

\begin{figure*}[!htbp]
\begin{center}
{\includegraphics[width=\columnwidth]{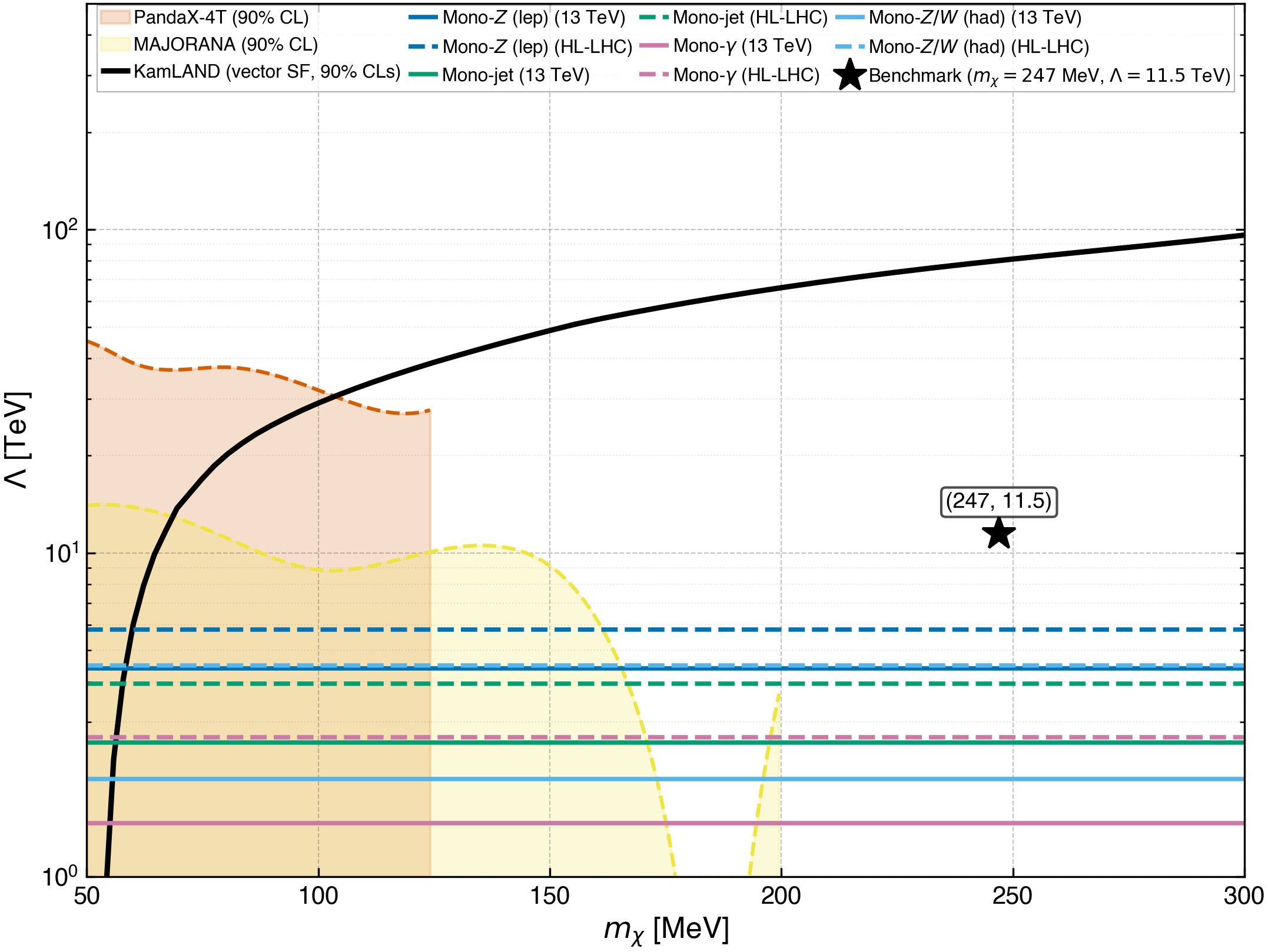}}
{\includegraphics[width=\columnwidth]{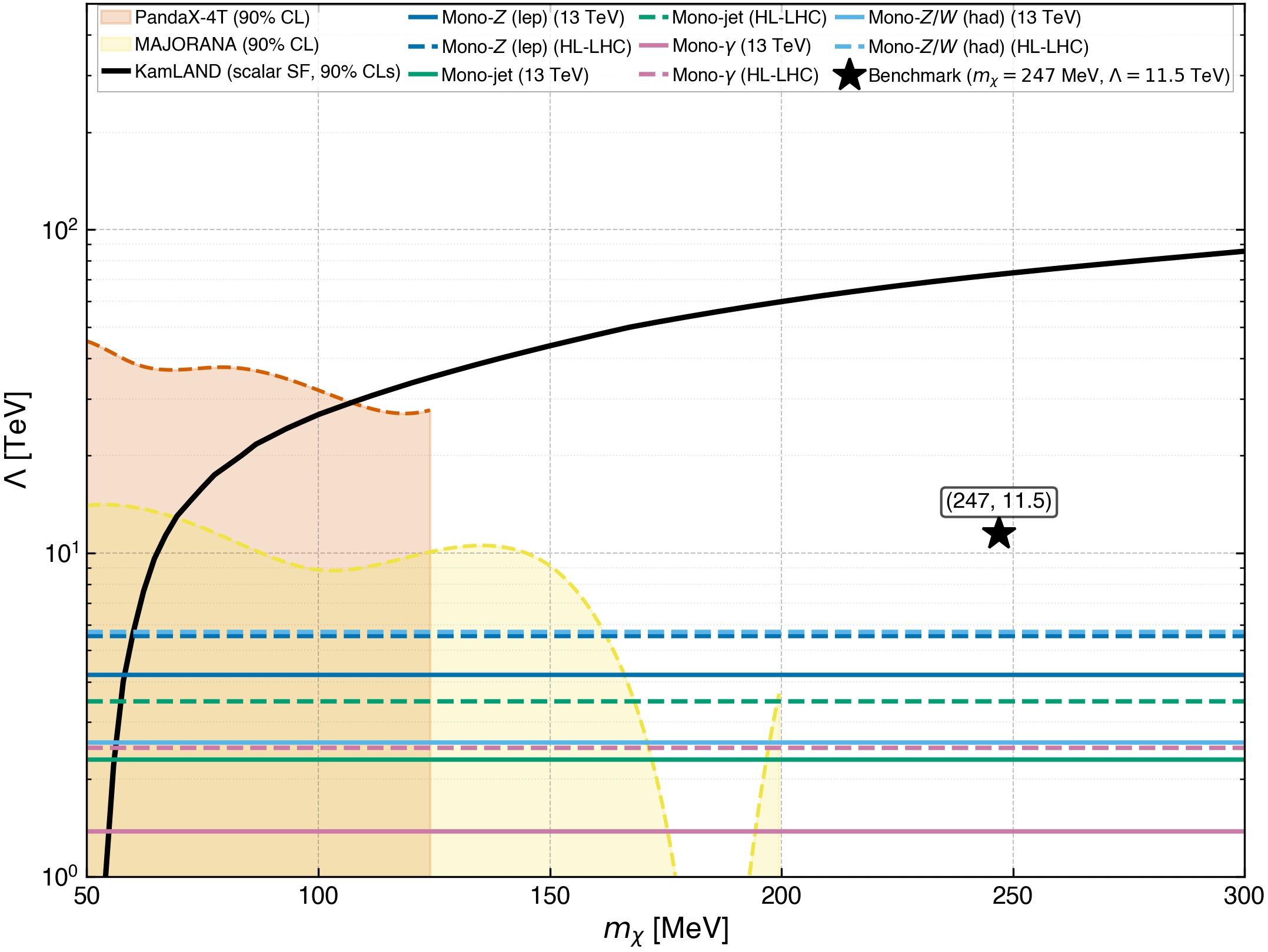}}
\caption{Exclusion regions for the vector (left) and scalar (right) operators. Solid curves show current bounds; dashed curves indicate projected sensitivities.}
\label{fig:constraints}
\end{center}
\end{figure*}

We now summarize existing constraints on neutral-current absorption of FDM, $\chi + (A,Z) \to \nu + (A,Z)$, focusing on their relevance to the $m_\chi \simeq 247\ \text{MeV}$ benchmark. The exclusion regions for the dark matter mass $m_{\chi}$ with the cutoff scale $\Lambda$ for both vector (left) and scalar (right) operators are shown in Figure~\ref{fig:constraints}. Each type of constraints are discussed below.  

\textbf{Collider constraints:} The absorption operators also induce mono-$X$ signatures at hadron colliders via associated $\chi\bar{\nu}$ production with initial-state radiation~\cite{Ma:2024tkt}. Using the ATLAS mono-$\gamma$ search with $139\ \text{fb}^{-1}$ at $\sqrt{s}=13\ \text{TeV}$, the 95\% C.L. lower limits on the cutoff scale $\Lambda$ for vanishing dark fermion mass are $\Lambda_V > 1.46\ \text{TeV}$ and $\Lambda_S > 1.38\ \text{TeV}$. The monojet channel provides stronger constraints: $\Lambda_V > 2.60\ \text{TeV}$, $\Lambda_S > 2.30\ \text{TeV}$. The most stringent limits come from mono-$Z$ with leptonic $Z$ decays: $\Lambda_V > 4.40\ \text{TeV}$, $\Lambda_S > 4.20\ \text{TeV}$. For the hadronic $Z/W$ channel with $36\ \text{fb}^{-1}$, the limits are $\Lambda_V > 2.00\ \text{TeV}$ and $\Lambda_S > 2.60\ \text{TeV}$. 

Projected sensitivities at the High-Luminosity LHC ($\sqrt{s}=14\ \text{TeV}$, $3\ \text{ab}^{-1}$) are: mono-$\gamma$: $\Lambda_V > 2.70\ \text{TeV}$, $\Lambda_S > 2.50\ \text{TeV}$; monojet: $\Lambda_V > 3.95\ \text{TeV}$, $\Lambda_S > 3.48\ \text{TeV}$; mono-$Z$ (leptonic): $\Lambda_V > 5.80\ \text{TeV}$, $\Lambda_S > 5.55\ \text{TeV}$; mono-$Z/W$ (hadronic): $\Lambda_V > 4.50\ \text{TeV}$, $\Lambda_S > 5.70\ \text{TeV}$~\cite{Ma:2024tkt}. 

\textbf{Direct detection:} PandaX-4T performed the first dedicated search for fermionic absorption using $95.0$ days of commissioning data, scanning $m_\chi$ from $30$ to $125\ \text{MeV}$~\cite{PandaX:2022osq}. No significant signal was found, with 90\% C.L. upper limits on the DM-nucleon absorption cross section ranging from $10^{-50}$ to $10^{-48}\ \text{cm}^2$, and the strongest limit $\sigma_{\chi N}^{\text{NC}} < 1.5\times10^{-50}\ \text{cm}^2$ at $m_\chi = 40\ \text{MeV}$. For our benchmark $m_\chi \simeq 247\ \text{MeV}$, the recoil energy $E_R \simeq 250\ \text{keV}$ lies above the PandaX-4T analysis window, so this search does not directly constrain the relevant parameter space.

The MAJORANA Demonstrator, using high-purity germanium detectors with a $37.5\ \text{kg}\cdot\text{yr}$ exposure, has set limits on fermionic absorption in the $1$--$100\ \text{keV}$ low-energy region~\cite{Majorana:2022gtu}. For $m_\chi$ from $25$ to $200\ \text{MeV}$, the cross-section bounds range from $10^{-47}$ to $10^{-43}\ \text{cm}^2$. The Helm form factor for germanium produces a pole-like structure at $m_\chi \simeq 174\ \text{MeV}$. For $m_\chi > 120\ \text{MeV}$, MAJORANA provides first bounds, though PandaX-4T surpasses these due to larger exposure.

The PICO-60 $\mathrm{C_3F_8}$ bubble chamber has set leading constraints on spin-independent absorption for $m_\chi < 23\ \text{MeV}$ and first limits on spin-dependent absorptive interactions~\cite{PICO:2025rku}. 

\textbf{Exotic meson decays:} Rare decays of $B$ and $K$ mesons can also constrain the vector operator through the $b\to s \chi \bar\nu$ or $s\to d \chi \bar\nu$ transitions. Recent analyses find that the resulting limits on $\Lambda_V$ are typically below $O(1)\ \text{TeV}$ and are subdominant compared to the LHC mono-$Z$ bounds~\cite{Feng:2026jsi}. We therefore neglect them in the following.

\textbf{Super-Kamiokande:} We assess whether existing Super-K data can probe the LZ benchmark for incoherent absorption. The signal involves two possible signatures, both incompatible with current public analyses. First, the knocked-out nucleon has kinetic energy $10$--$80\ \text{MeV}$, far below the Cherenkov threshold for protons in water, so the prompt–delayed strategy used in liquid scintillators is inapplicable. Second, the alternative $6.18\ \text{MeV}$ de-excitation $\gamma$ lies below the typical $7.5\ \text{MeV}$ prompt window of Super-K's atmospheric NCQE and DSNB searches, while higher-lying states lack reliable $\gamma$-branching inputs. We therefore conclude that Super-K cannot set robust limits with existing public data. A dedicated low-threshold SK-Gd analysis would be more promising; otherwise, other experiments or channels are required.

\textbf{Dark matter stability.} Fermionic absorption DM is intrinsically unstable. For the scalar portal, $\chi\to\nu\gamma\gamma$ is constrained by diffuse X-rays requiring $\tau_\chi\gtrsim10^{27}\s$; stellar cooling excludes $\sin^2\theta_\phi\gtrsim10^{-14}$ for $m_\phi\lesssim300\MeV$ in analogous models~\cite{Cox:2023cjw}. For the vector portal, $\chi\to3\nu$ and $\chi\to\nu\gamma\gamma\gamma$ set bounds~\cite{Dror:2019dib}, while the WZW channel requires dedicated matching.

\section{Conclusions}
\label{sec:conclusions}

We have shown that neutral-current absorption of fermionic dark matter on xenon nuclei provides a kinematically compelling explanation for the LZ230616 candidate event. For $m_\chi \simeq 247\ \mathrm{MeV}$, coherent absorption yields a monoenergetic recoil at $E_R \simeq 248\ \mathrm{keV}_{\mathrm{nr}}$ with cross section $\sigma_{\chi N}^{\mathrm{NC}} = 1.07\times10^{-46}\ \mathrm{cm}^2$ ($\Lambda \simeq 11.5\ \mathrm{TeV}$), producing one event in the LZ high-energy window. The same interaction inevitably produces $\mathcal{O}(10^2)$ primary incoherent knockouts with $T_{N'} \sim 10$--$100\ \mathrm{MeV}$, which are expected to be rejected by standard LZ selection cuts.

However, a recasting analysis of KamLAND data on the neutron-emission channel $\chi+{}^{12}\mathrm{C} \to \nu + n + {}^{11}\mathrm{C}^*$ excludes the benchmark parameter space required for the LZ interpretation. The KamLAND $90\%$ C.L. upper limit on the cross section lies below $1.07\times10^{-46}\ \mathrm{cm}^2$, establishing significant tension between the LZ excess and existing constraints from large-volume scintillator detectors.

While fermionic dark matter absorption remains a viable framework kinematically, the KamLAND constraint presented here indicates that the required parameter space is strongly challenged. Future dedicated searches for incoherent knockouts in LZ, improved nuclear spectral-function calculations, and independent constraints from JUNO and Borexino will be essential to determine whether this scenario can survive as an explanation for the LZ230616 event.

\begin{acknowledgments}
We thank Lei Wu and Yuanlin Gong for helpful discussions. This work is supported by the National Natural Science Foundation of China (NNSFC) under grants No. 12335005, No. 12575118 and the Special funds for postdoctoral overseas recruitment, Ministry of Education of China.
\end{acknowledgments}


\clearpage
\onecolumngrid

\section*{Supplemental Material}

\subsection*{S1. EFT Matching and the Helm Form Factor}

At quark level, the operators are
\begin{align}
\mathcal{O}_{\rm NC}^V&=\frac{1}{\Lambda_{\rm UV}^2}\sum_{q=u,d}C_q^V(\bar q\gamma^\mu q)(\bar\chi\gamma_\mu P_R\nu_R)+\mathrm{H.c.},\\
\mathcal{O}_{\rm NC}^S&=\frac{1}{\Lambda_{\rm UV}^2}\sum_{q=u,d}C_q^S(\bar qq)(\bar\chi P_R\nu_R)+\mathrm{H.c.}
\end{align}
Vector nucleon form factors at $q^2=0$: $F_{1u}^p=2,\ F_{1d}^p=1,\ F_{1u}^n=1,\ F_{1d}^n=2$. With $C_u^V=C_d^V$, $1/\Lambda_V^2=3C_q^V/\Lambda_{\rm UV}^2$. Scalar: $\langle N|\bar qq|N\rangle=(m_N/m_q)f_{Tq}^N\bar u_Nu_N$, so $g_S^N=\sum_q C_q^S f_{Tq}^N m_N/m_q$. 
The Helm form factor: 
\begin{equation}
    F_{H,j}(q)=3[\sin(qr_n)-qr_n\cos(qr_n)]/(qr_n)^3\exp(-q^2s^2/2) 
\end{equation}
with $r_n=1.14A_j^{1/3}\rm fm$, $s=0.9\rm fm$.

\subsection*{S2. Vector Portal UV Completion}

\subsubsection*{A. Anomaly Cancellation}
The SM baryon current is anomalous. The spectator set (in left-handed Weyl notation) is:
\begin{align}
\Psi_L&\sim(1,2,-1/2,y),& \Psi_R&\sim(1,2,-1/2,y+3), \notag \\
\eta_R&\sim(1,1,-1,y),& \eta_L&\sim(1,1,-1,y+3), \notag \\
\xi_R&\sim(1,1,0,y),& \xi_L&\sim(1,1,0,y+3). \notag
\end{align}
Taking $y=-3/2$ and a singlet $\zeta\sim(1,1,0,3)$ with $\vev{\zeta}$ generates spectator masses $M_\Psi\propto\vev{\zeta}$, $M_\eta\propto\vev{\zeta}$, $M_\xi\propto\vev{\zeta}$, ensuring no stable charged relics.

\subsubsection*{B. Mass Mixing and Matching}
The renormalizable Lagrangian is
\begin{equation}
\mathcal{L}_V=\mathcal{L}_{\rm SM}-\tfrac14V_{\mu\nu}V^{\mu\nu}+\frac{\epsilon_B}{2}V_{\mu\nu}B_Y^{\mu\nu}+|D_\mu\Phi_B|^2-V(\Phi_B)+\bar{\chi}(i\not D-m_{\chi0})\chi-(y_{\chi\nu}\Phi_B\bar{\chi}_L\nu_R+\mathrm{H.c.}). 
\end{equation}
After $\vev{\Phi_B}=v_B/\sqrt2$, $m_\chi=\sqrt{m_{\chi0}^2+\delta^2}$ with $\delta=y_{\chi\nu}v_B/\sqrt2$, $\sin\theta\equiv s_\theta=\delta/m_\chi$, $\cos\theta\equiv c_\theta=m_{\chi0}/m_\chi$. The $U(1)_B$ current gives $\mathcal{L}\supset-g_Bs_\theta c_\theta V_\mu(\bar\chi\gamma^\mu P_R\nu_R+\mathrm{H.c.})$. Integrating out $V$ with $m_V=g_Bv_B$ yields the matching relation.

Kinetic mixing $\epsilon_B$ induces leptonic couplings; $\epsilon_B$ must be tuned to suppress $\chi\to\nu_R e^+e^-, \nu_R \mu^+\mu^-$ ($m_\chi>2m_{e,\mu}$). The exact bound depends on the UV threshold.

\subsubsection*{C. WZW Decay}
For $m_{\pi^0}<m_\chi<2m_{\pi^0}$, the anomaly term is
\begin{equation}
\mathcal{L}_{\pi\gamma V}=c_{\rm WZW}\frac{eg_B}{16\pi^2F_\pi}\pi^0F_{\mu\nu}\widetilde V^{\mu\nu}, 
\end{equation} 
where $F_{\pi}\simeq 92.4$ MeV is the pion decay constant. 
The decay width is
\begin{equation}
\Gamma(\chi\to\nu_R\pi^0\gamma)=\frac{c_{\rm WZW}^2\alpha}{\Lambda_V^4F_\pi^2}m_\chi^7\mathcal P_3(m_{\pi^0}/m_\chi),
\end{equation}
where $\mathcal P_3$ is the dimensionless three-body phase integral. An order-one coefficient typically yields $\tau_\chi\ll10^{26}\s$ at $\Lambda_V = \mathcal{O}(10)\TeV$; precise exclusion requires complete anomaly matching.

\subsection*{S3. Scalar Portal UV Completion}

\subsubsection*{A. Gluonic Matching}
The scalar sector Lagrangian is
\begin{equation}
\mathcal{L}_S=\mathcal{L}_{\rm SM}+i\bar{\chi}\not\partial\chi-m_\chi\bar{\chi}\chi+i\bar{\nu}_R\not\partial\nu_R+\tfrac12(\partial\phi)^2-\tfrac12m_\phi^2\phi^2-(y_{\chi\nu}\phi\bar{\chi}_L\nu_R+\mathrm{H.c.}),
\end{equation}
with $\mathcal{L}_{\phi g}=\frac{\alpha_s}{12\pi}\frac{\kappa_g}{v}\phi G^2$. Using the QCD trace anomaly $\theta^\mu_{\ \mu}=\sum_q m_q\bar qq-\frac{9\alpha_s}{8\pi}G^2$, and the nucleon matrix element, we obtain $g_S^N=-m_N f_{TG}^N/(18v)$, $f_{TG}^N=1-\sum_q f_{Tq}^N$. Thus $\Lambda_S^{-2}=y_{\chi\nu}\kappa_g/m_\phi^2$. 

\subsubsection*{B. Radiative Decay and Meson Constraints}
Since $m_\phi>m_\chi$, $\chi\to\nu_R\phi$ is closed. Parity conservation forbids $\chi\to\nu_R\pi^0$ (because $\langle0|G^2|\pi^0\rangle=0$). The leading decay is $\chi\to\nu_R\gamma\gamma$, with amplitude $\mathcal F_{gg\gamma\gamma}(s)=\langle\gamma\gamma|G^2|0\rangle$. Because $s<m_\chi^2<4m_{\pi^0}^2$, no two-pion cut exists; a dispersive/lattice calculation is required. The lifetime must satisfy $\Gamma_{\chi,\rm tot}<6.58\times10^{-51}\GeV$ ($\tau_\chi>10^{26}\s$). On-shell $K^+\to\pi^+\phi$ is closed for $m_\phi>354\MeV$; off-shell $K^+\to\pi^+\chi\nu_R$ is suppressed and must be evaluated per model.

\subsection*{S4. Forbidden Freeze-Out and Cosmology}

The CP-odd singlet $a$ has $\mathcal{L}_{a\chi}=-ig_a a\bar\chi\gamma^5\chi$, $m_a>m_\chi$. The annihilation cross section is
\begin{equation}
\sigma(s)=\frac{\beta_a(s)}{32\pi s\beta_\chi(s)}\int_{-1}^{1}d\cos\theta\,|\mathcal M|^2,
\end{equation}
with $\beta_{a,\chi}=\sqrt{1-4m^2_{a,\chi}/s}$ and $\mathcal M$ given by $t$- and $u$-channel exchange. Thermal average:
\begin{equation}
\langle\sigma v\rangle(T)=\frac{1}{8m_\chi^4TK_2^2(m_\chi/T)}\int_{4m_a^2}^{\infty}ds(s-4m_\chi^2)\sqrt{s}K_1(\sqrt{s}/T)\sigma(s).
\end{equation} 
$K_1$, $K_2$ are modified Bessel functions of the second kind. 
Yield evolution:
\begin{equation}
\frac{dY_\chi}{dx}=-\frac{s}{Hx}\langle\sigma v\rangle(Y_\chi^2-Y_{\chi,\rm eq}^2),\quad x=m_\chi/T, 
\end{equation}
Here $s$, $H$ are entropy density and Hubble parameter, respectively. The observed relic $\Omega_\chi h^2=0.12$ fixes $g_a\sim\mathcal O(0.1-1)$ for $m_a/m_\chi\sim1.1-1.2$. The pseudoscalar $a$ must decay before BBN via a small Higgs-portal mixing.

Right-handed neutrinos contribute to $\Delta N_{\rm eff}$ if thermalized; decoupling follows $\Gamma_{\nu_R}(T_{\rm dec})=H(T_{\rm dec})$, with $\Gamma_{\nu_R}\sim C g_B^4s_\theta^4T^5/m_V^4$ in the vector portal.

\subsection*{S5. Microscopic matrix elements for incoherent absorption}

For the dark matter absorption process \(\chi(p_\chi) + N(p_N) \to \nu(p_\nu) + N'(p_{N'})\), where \(p_i = (E_i, \mathbf{p}_i)\) are four-momenta, the exact spin-averaged matrix elements squared are derived as follows.

For the vector operator, $\mathcal{O}_V = \frac{1}{\Lambda^2} (\bar{\nu} \gamma^\mu P_L \chi)(\bar{N}' \gamma_\mu N)$, the exact squared matrix element is
\begin{equation}
|\mathcal{M}_V|^2 = \frac{4}{\Lambda^4}
\left[
(p_{N'} \cdot p_\nu)(p_N \cdot p_\chi)
+ (p_{N'} \cdot p_\chi)(p_N \cdot p_\nu)
- m_N^2 (p_\nu \cdot p_\chi)
\right],
\end{equation} 
where \(a \cdot b = a_0 b_0 - \mathbf{a} \cdot \mathbf{b}\). For a stationary initial nucleon, this reduces to the leading-order expression \(|\mathcal{M}_V|^2 \to 4m_N^2 m_\chi^2/\Lambda^4\).

For the scalar operator, $\mathcal{O}_S = \frac{g_S}{\Lambda^2} (\bar{\nu} P_L \chi)(\bar{N}' N)$, the exact squared matrix element is: 
\begin{equation}
|\mathcal{M}_S|^2 = \frac{g_S^2}{\Lambda^4}
\cdot 2 \left( p_{N'} \cdot p_N + m_N^2 \right) (p_\nu \cdot p_\chi), 
\end{equation} 
with \(g_S\) set to 1 in this analysis. For a stationary initial nucleon, this reduces to:
\begin{equation}
|\mathcal{M}_S|^2 = \frac{4 g_S^2 m_N^2 m_\chi^2}{\Lambda^4}
\left( 1 + \frac{T_{N'}}{2m_N} \right)
\left( 1 - \frac{T_{N'}}{m_\chi} \right),
\end{equation} 
which shows the energy-dependent suppression of the scalar channel relative to the vector channel.

\subsection*{S6. Recasting analysis of KamLAND data}
The numerical rate calculation first evaluates these
eight bin counts at a fixed reference cross section
$\sigma_0^{\rm ref}$.  We denote the resulting array by
\begin{equation}
R_i^{\rm ref}
\equiv
s_i(\sigma_0^{\rm ref}).
\end{equation}
Since the event rate is linear in $\sigma_0$, we rescale this template
before the statistical scan according to
\begin{equation}
\sigma_0^{\rm template}
=
\sigma_0^{\rm ref}
\frac{10}{\max_i R_i^{\rm ref}},
\qquad
s_i^{\rm template}
=
R_i^{\rm ref}
\frac{\sigma_0^{\rm template}}{\sigma_0^{\rm ref}}.
\label{eq:kamland-template}
\end{equation}
This numerical rescaling ensures that the largest reference signal bin
contains ten events.  The signal strength parameter $\mu$ then gives
\begin{equation}
s_i(\mu)=\mu s_i^{\rm template},
\qquad
\sigma_0=\mu\sigma_0^{\rm template}.
\label{eq:kamland-mu-sigma}
\end{equation}

The observed counts, background central values, and adopted $25\%$
background uncertainty follow
Refs.~\cite{Gong:2025ves,MeighenBerger:2023tad}. We use the eight observed event counts
\begin{equation}
n_i=(4,\,2,\,2,\,2,\,2,\,1,\,1,\,1)
\end{equation}
and background central values
\begin{equation}
b_i=(3.9,\,3.1,\,2.9,\,2.2,\,2.1,\,2.0,\,1.9,\,1.9).
\label{eq:kamland-backgrounds}
\end{equation}
Each background bin is assigned an independent $25\%$ uncertainty,
\begin{equation}
\delta b_i=0.25\,b_i,
\qquad
\tau_i=
\left(\frac{b_i}{\delta b_i}\right)^2=16.
\label{eq:kamland-background-uncertainty}
\end{equation}
Here $\tau_i$ specifies the auxiliary Poisson constraint used to model
the independent background uncertainty in each bin.

We introduce one background nuisance parameter $\gamma_i$ per bin.  The
predicted count is
\begin{equation}
\lambda_i(\mu,\gamma_i)
=
\mu s_i^{\rm template}+\gamma_i b_i,
\end{equation}
where $\gamma_i=1$ corresponds to the nominal background.  The
likelihood is
\begin{align}
\mathcal L(\mu,\boldsymbol{\gamma})
={}&
\prod_{i=1}^{8}
\operatorname{Pois}
\left(
n_i\mid
\mu s_i^{\rm template}+\gamma_i b_i
\right)
\nonumber\\
&\times
\prod_{i=1}^{8}
\operatorname{Pois}
\left(
\tau_i\mid
\gamma_i\tau_i
\right).
\label{eq:kamland-likelihood}
\end{align}
The second product implements the assumed bin by bin background
uncertainty and does not represent additional KamLAND events.

For each value of $m_\chi$, we evaluate the profile likelihood $CL_s$
procedure separately for the vector and scalar operators,
$\mathcal O=V,S$~\cite{Read:2002hq,Cowan:2010js,Heinrich:2021zdu}.
The observed $90\%$ confidence level limit is defined by
\begin{equation}
CL_s(\mu_{90})=0.1,
\qquad
\sigma_{0,90}^{\mathcal O}
=
\mu_{90}\sigma_0^{\rm template}.
\label{eq:kamland-sigma-limit}
\end{equation}
We report this result as a lower bound on the interaction scale using
the convention
\begin{equation}
\sigma_0=
\frac{m_\chi^2}{4\pi\Lambda^4},
\qquad
\Lambda_{90}^{\mathcal O}
=
\left(
\frac{m_\chi^2}
{4\pi\sigma_{0,90}^{\mathcal O}}
\right)^{1/4}.
\label{eq:kamland-sigma-lambda}
\end{equation}
\clearpage
\twocolumngrid

\bibliography{refs}
\bibliographystyle{JHEP}
\end{document}